\documentstyle[12pt,epsf]{article}
\newcommand{\beq}{\begin{equation}}
\newcommand{\eeq}{\end{equation}}
\begin{document}
\begin{titlepage}
%\title{\bf Higgs Mass Bound in the Minimal Standard Model}
\title{\bf Higgs Mass Bound in the Minimal Standard Model\thanks{
Invited talk given at the 4th Hellenic School on Elementary Particle
Physics, Corfu, Greece, Sept. 2--20, 1992. To appear in the proceedings.}}
{\bf
\author{
   Urs M. Heller \\
   Supercomputer Computations Research Institute \\
   Florida State University \\
   Tallahassee, FL 32306 \\
%   U.S.A.}
   U.S.A. \\[+1.5cm]
   FSU-SCRI-93-40}
}
\date{March 1993}
%\date{}
\maketitle
\begin{abstract}
\noindent
A brief review of the role of the Higgs mechanism and the ensuing Higgs
particle in the Minimal Standard Model is given. Then the property of
triviality of the scalar sector in the Minimal Standard Model and the upper
bound on the Higgs mass that follows is discussed. It is emphasized that
the bound is obtained by limiting cutoff effects on physical processes.
Actions that allow a parameterization and tuning of the leading cutoff
effects are studied both analytically, in the large $N$ limit of the
generalization of the $O(4)$ symmetry of the scalar sector to $O(N)$, and
numerically for the physical case $N = 4$. Combining those results
we show that the Minimal Standard Model will describe physics to
an accuracy of a few percent up to energies of the order 2 to 4 times
the Higgs mass, $M_H$, only if $M_H \le 710 \pm 60 ~ GeV$. This bound is
the result of a systematic search in the space of dimension six
operators and is expected to hold in the {\it continuum}.
\end{abstract}
\thispagestyle{empty}
\end{titlepage}
\pagestyle{empty}

{\Large{\bf 1. Introduction}}
\vspace{.5cm}

The elementary particles and their interactions are described in a highly
economical and successful way in the Minimal Standard Model. The symmetry
among the elementary fermions is described by the internal symmetry group
\beq
G = SU(3)_{\rm color} \times SU(2)_{\rm L} \times U(1)_{\rm Y}
\eeq
and we have three known families of such fermions. Indeed, precision
measurements of the width of the $Z$ vector boson at LEP have yielded for the
number of massless neutrinos (and thus of families, since we have one
massless lefthanded neutrino per family) $N_\nu = 3.04 \pm 0.04$
\cite{neutrino}.

The interactions among the elementary fermions are introduced by making the
global symmetry group $G$ into a local symmetry group with the help of $8 +
3 + 1 = 12$ massless vector bosons ($8$ gluons for the $SU(3)_{\rm color}$
factor, 3 vector bosons for $SU(2)_{\rm L}$ and 1 for the $U(1)$
hypercharge). The resulting theory is perturbatively renormalizable, which
means in particular that it is applicable for energies ranging over many
orders of magnitude and that it has a lot of predictive power.

However, so far we do not describe nature appropriately. We know that
explicitly realized is only the subgroup $SU(3)_{\rm color} \times
U(1)_{\rm em}$ of $G$, where here the $U(1)$ factor describes
electromagnetism. Furthermore the theory does not allow for masses. In the
case of fermions, mass terms are forbidden by the chiral nature of the
symmetry group $G$, and in the case of the vector bosons, mass terms are
forbidden by the gauge symmetry. But in nature 14 out of the 15 fermions
per family are massive, as are the weak vector bosons $W^{\pm}$ and $Z$. In
the case of the three vector bosons we don't even have enough degrees of
freedom, since massless vector boson have two (transverse) polarizations,
while massive ones need in addition a state with longitudinal polarization.

In the Minimal Standard Model we correct these shortcomings by adding an
elementary complex scalar field, transforming as a doublet under
$SU(2)_{\rm L}$. Group theory determines that the scalar self-interactions
have an enhanced symmetry, $SU(2)_{\rm L} \times SU(2)_{\rm custodial}
\simeq O(4)$. This symmetry is then arranged to be broken spontaneously to
$O(3)$ by giving the scalar field a non-vanishing vacuum expectation value,
$F$. (This notation is chosen from the analogy to the current algebra
describing the soft pions of QCD, in which the analog of the vacuum
expectation value is the pion decay constant, $f_\pi$). This spontaneous
symmetry breaking turns out to do all the tricks we need. It breaks the
symmetry group from $G$ to the subgroup $SU(3)_{\rm color} \times U(1)_{\rm
em}$, explicitly realized. The three Goldstone bosons of the breaking $O(4)
\to O(3)$ provide the longitudinal degrees of freedom of the $W^{\pm}$ and
$Z$, making those massive, and finally the spontaneous symmetry breaking
gives masses to the fermions via gauge invariant Yukawa couplings. And all
these good things happen while the theory remains perturbatively
renormalizable and maintains its predictive power. After the spontaneous
symmetry breaking, one out of the 4 scalar degrees of freedom that we added
is left, the so far elusive Higgs boson.

While the Minimal Standard Model, briefly outlined above, has had many
spectacular successes, relatively little is known experimentally about the
Higgs sector. One exception is the value of the vacuum expectation value
$F$. From its relation to the $W$ boson mass and the latter's to the
four-fermion coupling $G_F$ one can easily deduce that $F = 246 ~ GeV$.
Furthermore, recent experiments at LEP led to a lower bound on the Higgs
mass of about $60 ~ GeV$ \cite{lowerbound}.

In the remainder of this seminar I will describe what we know about the
mass of this Higgs particle on purely theoretical grounds. In particular I
will present the arguments leading to an upper bound on the mass of the
Higgs particle, the so called triviality bound. And finally I'll describe a
non-perturbative computation of the bound, that we found to be $710 \pm 60
{}~ GeV$.

\vspace{1cm}
{\Large{\bf 2. Perturbative indications of triviality}}
\vspace{.5cm}

As indicated in the introduction we are interested in an upper bound on the
Higgs mass. It turns out, as we will see below, that the Higgs mass
increases with increasing scalar self coupling. At the energy scale of the
upper bound of Higgs mass, all gauge interactions are relatively weak and
can be treated perturbatively. The same holds true for the Yukawa
interactions, including the top if it is not heavier than about $200 ~
GeV$, as is favored by experiment \cite{neutrino}. Therefore we concentrate
here on the scalar sector of the Minimal Standard Model alone. It is
described by an $O(4)$ invariant scalar field theory with potential
\beq
V(\vec \phi) = \frac{1}{2} \mu^2_0 \vec \phi^2 + \frac{g_0}{4!} (\vec
\phi^2)^2
\eeq
To have spontaneous symmetry breaking we assume $\mu^2_0 < 0$. As usual the
theory so far described is ill defined. We need to introduce a cutoff
$\Lambda$ to regulate and then renormalize the theory. A two-loop
perturbative computation together with application of the renormalization
group leads to the relation between the cutoff, $\Lambda$, the physical
Higgs mass, $M_H$, and the renormalized coupling, $g_R = 3M^2_H/F^2$
\beq
\frac{M_H}{\Lambda} = C \left( \frac{g_R}{4\pi^2} \right)^{13/24} \exp
\left\{ -\frac{4\pi^2}{g_R} \right\} ~ \left[ 1 + O(g_R) \right]
\label{eq:MoverL}
\eeq
Usually, at the end of the calculation one would like to remove the cutoff
by taking the limit $\Lambda \to \infty$. However, from
eq.~(\ref{eq:MoverL}) we see that the limit $\Lambda \to \infty$ implies
$g_R \to 0$, {\it i.e.} we are left with a non-interacting, {\bf trivial}
theory. But we need an interacting scalar sector for the Higgs mechanism to
work. Therefore we need to keep the cutoff finite: the Minimal Standard
Model has to be viewed as an effective theory that describes physics at
energies below the cutoff scale.

Since we have to keep a finite cutoff, we may ask what happens if
we try to make the (renormalized) scalar self-interactions stronger.
Eq.~(\ref{eq:MoverL}) tells us that as $g_R$ increases, so does the ratio
$M_H/\Lambda$. But since the Higgs mass is one of the physical
quantities that the standard model is supposed to describe, we certainly
need $M_H/\Lambda < 1$. Hence we have arrived at an upper bound on
the Higgs mass. Since it comes from the triviality of the scalar sector,
this bound is referred to as the triviality bound.

In the next three sections I will make the definition of the bound, namely
the meaning of the ``$<$'' in $M_H/\Lambda < 1$ more precise and
give the results of a numerical computation of the bound.

\vspace{1cm}
{\Large{\bf 3. Cutoff effects and generalized actions}}
\vspace{.5cm}

The triviality of the scalar sector of the Minimal Standard Model, and
therefore the need to retain a finite cutoff $\Lambda$, are by now very well
established \cite{oldHiggs,LW,genHiggs,BBHN2}. As a consequence, all
observable predictions have a weak cutoff dependence, of order
$1/\Lambda^2$. I will later on show explicit examples of such
cutoff effects, computed in the large $N$ limit of the generalization of
the $O(4)$ symmetric scalar sector to $O(N)$. This generalization is
useful, because we can solve the model analytically in the large $N$ limit,
and therefore {\it e.g.} compute cutoff effects. The cutoff effects become
larger when the ratio $M_H/\Lambda$, and hence $g_R$, increases.
Therefore, by limiting the cutoff effect on some physical observable -- we
shall use the square of the invariant scattering amplitude of Goldstone
bosons at $90^o$ in the center of mass frame -- we obtain a more precise
definition of the upper bound on the Higgs mass.

The need for a finite cutoff in the Minimal Standard Model means, that it
is only an effective theory, applicable for energies below the cutoff.
There will be some, as yet unknown, embedding theory. But we assume that
for ``small" energies -- energies smaller than a few times the Higgs mass
-- the scalar sector is representable by an effective action (for a more
detailed discussion see {\it e.g.} \cite{Dallas} and references therein)
\beq
L_{\rm eff} = L_{\rm ren} + \frac{1}{\Lambda^2} \sum_A c_A O_A ~,
 ~~~~~~~ {\rm dim} O_A \le 6
\label{eq:effA}
\eeq
with $L_{\rm ren}$ the usual renormalized $\phi^4$ Lagrangian. $O_A$ are
operators with the correct symmetry properties and dimension less than or
equal to 6, and the coefficients $c_A$ depend on the embedding theory.
Since we don't know this theory, we just parameterize our ignorance with
these $c_A$'s, {\it i.e.,} we consider reasonable bare cutoff models with
enough free parameters to reproduce the effective action,
eq.~(\ref{eq:effA}). This allows us to tune the cutoff effects. Eliminating
redundant operators, which leave the S-matrix unchanged, we end up with two
``measurable" $c_A$'s. The Higgs mass bound is now obtained as the maximal
value $M_H$ can take, when varying the parameters $c_A$ while maintaining
our requirement of limiting the cutoff effects by some prescribed value,
typically a few percent. However, since we do not know the embedding
theory, we will in this maximization avoid excessive fine tuning that might
eliminate leading cutoff effects, of order $1/\Lambda^2$.

The most straightforward implementation is to start with an action of
the form eq.~(\ref{eq:effA}) on the level of bare fields and parameters.
It turns out, however, that, as our intuition would tell us, the maximal
renormalized coupling $g_R$, and hence the maximal Higgs mass, is obtained
at maximal bare $\phi^4$ self coupling, {\it i.e.,} at $g_0 \to \infty$ (see
{\it e.g.} \cite{LW,LrgN}). But in this limit the model becomes nonlinear,
with the field having a fixed length, and all dimension six operators
become trivial. In a non-linear theory it are terms with four derivatives
that allow us to tune the cutoff effects of order $1/\Lambda^2$.
Maintaining $O(N)$ invariance, there are three different terms with four
derivatives, and we are led to consider actions of the form
\beq
S ~ = \int_x ~ \left [\frac{1}{2} \vec \phi ( -\partial^2 + 2b_0 \partial^4)
\vec \phi - \frac{b_1}{2N} (\partial_\mu \vec \phi
\cdot \partial_\mu \vec \phi )^2 - \frac{b_2}{2N} (\partial_\mu \vec
\phi \cdot \partial_\nu \vec \phi - \frac{1}{4} \delta_{\mu , \nu }
\partial_\sigma \vec \phi \cdot \partial_\sigma \vec \phi )^2 \right]
\label{eq:S_c}
\eeq
with $\phi^2 = N \beta$ fixed. Up to terms with more derivatives, the
parameter $b_0$ can be eliminated with a field redefinition
\beq
\vec \phi \rightarrow \frac{\vec \phi + b_0 \partial^2 \vec \phi }
{\sqrt{\vec \phi^2 + b_0^2 (\partial^2 \vec \phi )^2 + 2b_0 \vec \phi
\partial^2 \vec \phi }} \sqrt{N\beta} .
\label{eq:field_red}
\eeq
This leaves two free parameters to tune the cutoff effects, exactly the
number of measurable coefficients $c_A$ in (\ref{eq:effA}). Therefore
(\ref{eq:S_c}) should give a good parameterization to obtain the Higgs mass
bound.

\vspace{1cm}
{\Large{\bf 4. Solution at large $N$}}
\vspace{.5cm}

To understand the effect of the four-derivative couplings in the action
eq.~(\ref{eq:S_c}) we studied these models first in the soluable large $N$
limit \cite{LrgN}. We considered different regularizations: a class of
Pauli-Villars regularizations obtained by replacing the term $\vec \phi (
-\partial^2) \vec \phi$ by $\vec \phi ( -\partial^2 ( 1 + (- \partial^2/
\Lambda^2)^n ) \vec \phi$ with $n \ge 3$ and $b_0$ set to zero, and some
transcriptions of action (\ref{eq:S_c}) on a lattice, {\it i.e.} lattice
actions such that eq.~(\ref{eq:S_c}) appears in their expansion in slowly
varying fields.

The result of our investigations is that at $N=\infty$, after $b_0$ has
been eliminated, $b_2$ has no effect and that the bound depends
monotonically on $b_1$, increasing with decreasing $b_1$. Overall stability
of the homogeneous broken phase restricts the range of $b_1$ and thus we
find a finite optimal value for $b_1$. The physical picture that emerges is
that among the nonlinear actions the bound is further increased by reducing
as much as possible the attraction between low momentum pions in the
$I=J=0$ channel.

The rule in the above paragraph does not lead to an exactly universal
bound. Different bare actions that give the same effective parameter
$b_1$ can give somewhat different bounds because the dependence of
physical observables on the bare action is highly nonlinear. For
example, at the optimal $b_1$ value, Pauli--Villars regularizations
lead to bounds higher by about $100 ~ GeV$ than some lattice
regularizations. This difference between the lattice and
Pauli--Villars can be traced to the way the free massless inverse
Euclidean propagator departs from the $O(p^2 )$ behavior at low
momenta. For Pauli--Villars it bends upwards to enforce the needed
suppression of higher modes in the functional integral, while on the
lattice it typically bends downwards to reflect the eventual
compactification of momentum space.

When considering lattice regularizations, because we desire to preserve
Lorentz invariance to order $1/\Lambda^2$, we use the $F_4$ lattice. The
$F_4$ lattice can be thought of as embedded in a hypercubic lattice from
which odd sites (i.e. sites whose integer coordinates add up to an odd sum)
have been removed. The $F_4$ lattice turns out to have a larger symmetry
group than the hypercubic lattice, which forbids Lorentz invariance
breaking terms at order $1/\Lambda^2$.

%Figure 1
\begin{figure}[htb]
%\vspace{12.cm}
%\epsfxsize=\columnwidth
\epsfxsize=12.cm
\epsffile{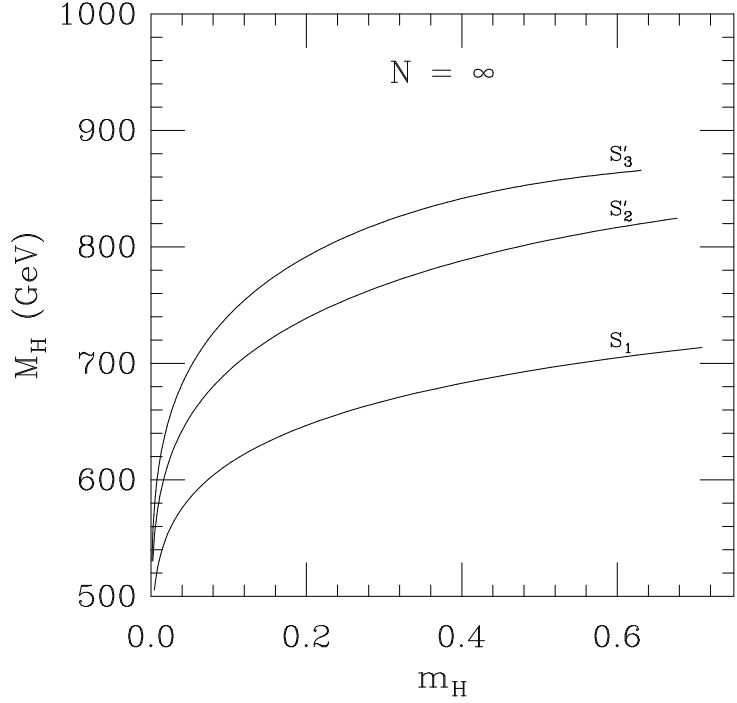}
\caption{Large $N$ prediction of the Higgs mass $M_H = M_H / F \times 246 ~
GeV$ in physical units vs. the Higgs mass $m_H = a M_H$ in lattice units
for the three actions on the $F_4$ lattice}
\label{fig:LrgN_m_f}
\end{figure}

On the basis of the above observations, we went through three stages of
investigation. The first stage was to investigate the na\"{\i}ve
nearest--neighbor model. This should be viewed as the generic lattice case
where no special effort to increase the bound is made. The next stage is to
write down the simplest action that has a tunable parameter $b_1$.  We
should emphasize that on the $F_4$ lattice, unlike on the hypercubic
lattice, this can be done in a way that maintains the nearest--neighbor
character of the action, namely by coupling fields sited at the vertices of
elementary bond--triangles. The last stage is to add a term to eliminate
the ``wrong sign'' order $p^4$ term in the free propagator, amounting to
Symanzik improvement. The three $F_4$ actions we investigated are given by
\begin{eqnarray}
S_1 = & -2\beta_0 \sum_{<x,x'>} \vec \Phi (x) \cdot \vec \Phi (x')
\nonumber \\
S_2 = & S_1 ~-~ \frac{\beta_2}{8} \sum_x
\sum_{{{<ll'>}\atop {l,l' \cap  x \ne
\emptyset ,~ l\cap x' \ne\emptyset ,~ l'\cap x'' \ne\emptyset }}\atop
{x,x',x'' {}~{\rm all~ n.n.}} } \left [ \left( \vec \Phi (x) \cdot \vec \Phi
(x') \right) ~ \left(\vec \Phi (x) \cdot \vec \Phi (x'') \right) \right ]
\label{eq:NumAct} \\
S_3 = & - {2(2\beta_0 +\beta_2 )}\sum_{<x,x'>}  \vec \Phi (x) \cdot \vec
\Phi (x') ~+ ~{(\beta_0 +\beta_2 )}\sum_{\ll x,x'\gg}  \vec \Phi (x) \cdot
\vec \Phi (x') \nonumber \\
& - \frac{\beta_2}{8} \sum_x \sum_{{{<ll'>}\atop {l,l' \cap  x \ne
\emptyset ,~ l\cap x' \ne\emptyset ,~ l'\cap x'' \ne\emptyset }}\atop
{x,x',x'' {}~{\rm all~ n.n.}} } \left [ \left( \vec \Phi (x) \cdot \vec \Phi
(x') \right) ~ \left(\vec \Phi (x) \cdot \vec \Phi (x'') \right) \right ] .
\nonumber
\end{eqnarray}
The coupling $\beta_2$ here plays the role of $b_1$ in eq.~(\ref{eq:S_c}).

In the large $N$ limit, to simplify the calculation somewhat, we actually
considered slight variants of actions $S_2$ and $S_3$ (see \cite{LrgN}),
that have the same expansion for slowly varying fields and are expected to
give the same physics. We indicate this by denoting the modified actions
with a ${}^\prime$. In each case, at constant $\beta_2$, $\beta_0$ is
varied tracing out a line in parameter space approaching a critical point
from the broken phase. This line can also be parameterized by $M_H/\Lambda$
(on the lattice $\Lambda = a^{-1}$) or $g_R$. For actions $S^\prime_2$ and
$S^\prime_3$, $\beta_2$ is chosen so that on this line the bound on $M_H$
is expected to be largest. A simulation produces a graph showing $M_H/F$ as
a function of $M_H/\Lambda$ along this line. The y-axis is turned into an
axis for $M_H$ by $M_H = M_H/F \times 246 ~ GeV$.  The large $N$
predictions for these graphs are shown in Figure~\ref{fig:LrgN_m_f}.

\begin{figure}[htb]
%\vspace{12.cm}
%\epsfxsize=\columnwidth
\epsfxsize=12.cm
\epsffile{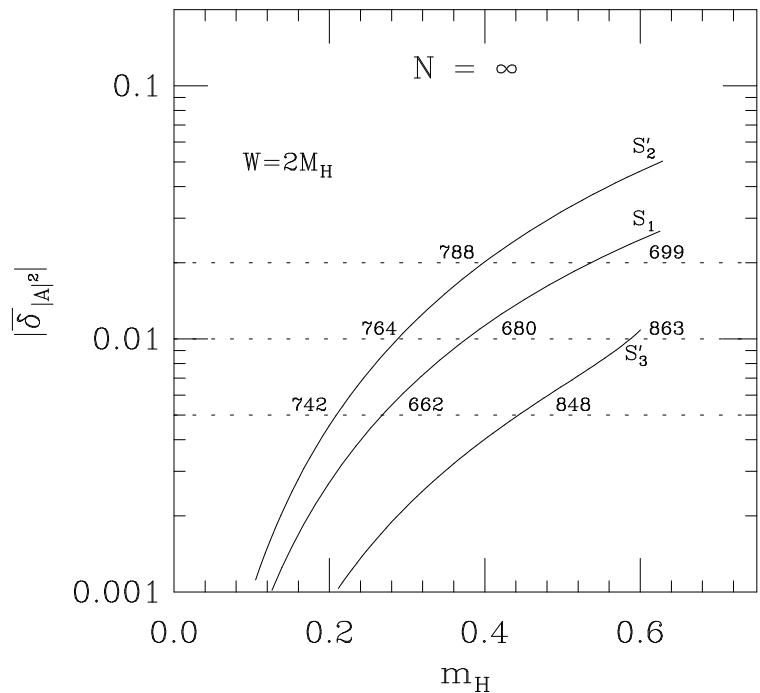}
\caption{Leading order cutoff effects in the invariant $\pi-\pi$ scattering
amplitude at $90^o$ at center of mass energy $W=2M_H$ vs. the Higgs mass
$m_H = a M_H$ in lattice units for the three actions. The values of $M_H$
in $GeV$ determined from $M_H = M_H / F \times ~ GeV$ are put on the three
horizontal lines at $\bar \delta_{|A|^2} = 0.005,~0.01,~0.02$.}
\label{fig:cutoff}
\end{figure}

To obtain a well defined bound on the Higgs mass from such graphs, as we
already explained, we need to compute the cutoff effects on some physical
quantity. We show the cutoff effect in the square of the invariant $\pi-\pi$
scattering amplitude at $90^o$ at several center of mass energies in
Figure~\ref{fig:cutoff}.

%Figure 2

If one considers only the magnitude of cutoff effects as a function of $M_H
/ \Lambda$, one might conclude that the bound obtained with action $S_1$
would be larger than the bound obtained with $S^\prime_2$. This conclusion
proves to be wrong when the mass in physical units is considered. The
values of $M_H$ in $GeV$, determined from $M_H = M_H / F \times 246 ~ GeV$,
are put on three horizontal lines in Figure~\ref{fig:cutoff} at $\bar
\delta_{|A|^2} = 0.005,~0.01,~0.02$. At large $N$ the bound increases when
going from $S_1$ to $S^\prime_2$ and then to $S^\prime_3$ by a little over
10\% at each step. For example, for $\bar\delta_{|A|^2} =.01$ we get bounds
on $M_H$ of 680, 764, 863 $GeV$ for $S_1$, $S^\prime_2$ and $S^\prime_3$
respectively.

\vspace{1cm}
{\Large{\bf 5. The physical case $N=4$.}}
\vspace{.5cm}

\begin{figure}[htb]
%\vspace{12.cm}
%\epsfxsize=\columnwidth
\epsfxsize=12.cm
\epsffile{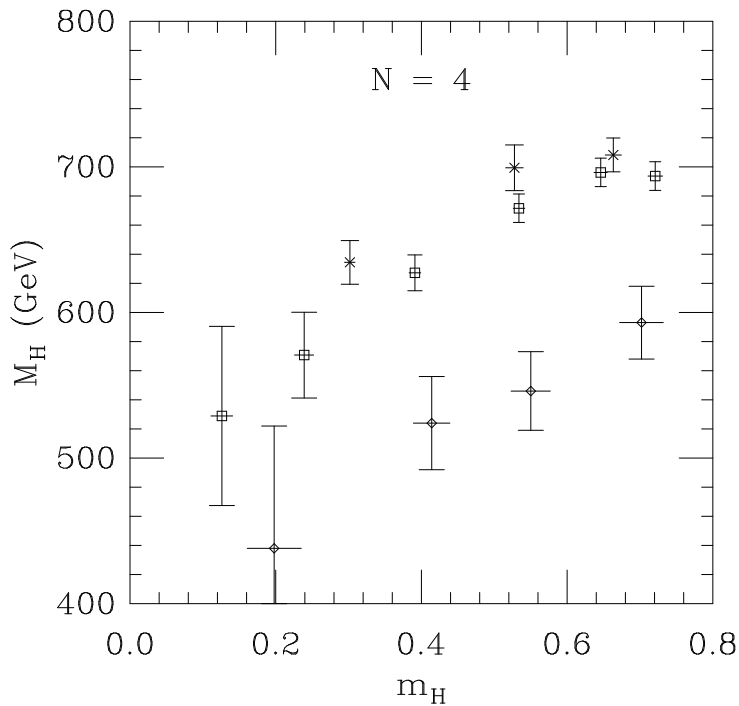}
\caption{The Higgs mass $M_H = M_H / F \times 246 ~ GeV$ in physical units
vs. the Higgs mass $m_H = a M_H$ in lattice units from the numerical
simulations. The diamonds correspond to action $S_1$ \protect \cite{BBHN2},
the squares to action $S_2$ and the crosses to action $S_3$ \protect
\cite{4der_num}.}
\label{fig:num_m_f}
\end{figure}

For the physical case, $N = 4$, we do not have at our disposal
non-perturbative analytical methods of computation. We therefore resort to
numerical simulations of the models in eq.~(\ref{eq:NumAct}).  The result
of these simulations are show in Figure~\ref{fig:num_m_f} which shows $M_H
=246\sqrt{g_R/3} ~GeV$ as a function of $a M_H$ for all three actions. One
clearly sees the progressive increase of $M_H$ from action $S_1$ to
$S_2$ and then $S_3$, just as in the large $N$ limit.

%Figure 3

To obtain a bound on the Higgs mass we also need estimates of the cutoff
effects. We do not know how to compute cutoff effects in numerical
simulations. Therefore we take the estimates from the large $N$
calculation, which should be accurate enough for our purpose here (observe
that the cutoff effects are relatively insensitive to the Higgs mass in
lattice units, $m_H = a M_H$). A glance at Figure~\ref{fig:cutoff} shows
that in all cases the cutoff effects on the pion--pion scattering are below
a few percent even at the maximal $M_H$ of each curve. Thus we can take the
largest of these maxima as our bound. The ordering of the points and their
relative positions are in agreement with Figure~\ref{fig:LrgN_m_f}, while
the differences in overall scale, reflecting the difference between
$N=\infty$ and $N=4$, come out compatible with $1/N$ corrections, as
expected \cite{LrgN}.

We conclude that the Minimal Standard Model will describe physics to an
accuracy of a few percent up to energies of the order 2 to 4 times the
Higgs mass, $M_H$, only if $M_H \le 710 \pm 60 ~ GeV$. The error quoted
accounts for the statistical errors, shown in Figure~\ref{fig:num_m_f}, as
well as the systematic uncertainty associated with the remaining
regularization dependence, {\it e.g.} a not completely optimal choice of
$b_1$ in (\ref{eq:S_c}) and the possible small dependence on $b_2$ for $N
\neq \infty$. Since this bound is the result of a systematic search in the
space of dimension six operators, we expect it to hold in the {\it
continuum}. A Higgs particle of mass $710 ~ GeV$ is expected to have a
width between $180 ~ GeV$ (the perturbative estimate) and $280 ~ GeV$ (the
large $N$ non-perturbative estimate). Thus, if the Higgs mass bound turns
out to be saturated in nature, the Higgs would be quite strongly
interacting.

\vspace{1cm}
{\Large{\bf Acknowledgements}}
\vspace{.5cm}

I would like to thank my collaborators, H. Neuberger, P. Vranas and M.
Klomfass for a most fruitful and stimulating collaboration, and the
organizers of the School for inviting me to contribute and for organizing a
very pleasant and interesting School. This work was supported in part by
the DOE under grants \# DE-FG05-85ER250000 and \# DE-FG05-92ER40742.

\vfill
\newpage
\vspace{.5cm}
%\begin{thebibliography}{99}

\vfill
\end{document}